\documentclass[12pt,a4paper]{article}
\usepackage[dvips]{graphicx,color}
\newcommand{\be}{\begin{equation}}
\newcommand{\ee}{\end{equation}}
\newcommand{\ba}{\begin{eqnarray}}
\newcommand{\ea}{\end{eqnarray}}
\newcommand{\baa}{\begin{eqnarray*}}
\newcommand{\eaa}{\end{eqnarray*}}

\begin{document}
{\pagestyle{empty}
\vskip 1.5cm

{\renewcommand{\thefootnote}{\fnsymbol{footnote}}
\centerline{\large \bf 
Optimization of protein force-field parameters with the Protein Data Bank}
}
%\vskip 3.0cm
\vskip 2.0cm

\centerline{Yoshitake Sakae$^{a}$
and Yuko Okamoto$^{a,b}$}
\vskip 1.5cm
\centerline{$^a${\it Department of Functional Molecular Science}}
\centerline{{\it The Graduate University for Advanced Studies}}
\centerline{{\it Okazaki, Aichi 444-8585, Japan}}
\centerline{and}
\centerline{$^b${\it Department of Theoretical Studies}}
\centerline{{\it Institute for Molecular Science}}
\centerline{{\it Okazaki, Aichi 444-8585, Japan}}

%\vskip 1.5cm

\medbreak
\vskip 1.0cm

\centerline{\bf Abstract}
\vskip 0.3cm

We propose a novel method to optimize existing force-field parameters
for protein systems.
The method consists of minimizing the summation
of the square of the force acting on each atom in the proteins
with the structures from the Protein Data Bank.
We performed this optimization to the partial-charge and 
torsion-energy parameters of
the AMBER parm96 force field, 
using 100 molecules from the Protein Data Bank.
We then performed folding
simulations of $\alpha$-helical and $\beta$-hairpin peptides.
The optimized force-field parameters
gave structures more similar
to the experimental implications
than the original AMBER force field.

%\vfill
%\newpage}
}
%\baselineskip=0.8cm

% main text
   
\section{Introduction}
\label{introduction}
In the theoretical studies of biomolecular systems, it is now
indispensable to use molecular simulation techniques such as
Monte Carlo (MC) and molecular dynamics (MD) methods.
Most such simulations use the potential energy functions within the 
framework of classical mechanics consisting of certain energy terms
with force-field parameters.
Well-known force fields (or potential energy functions) are
AMBER \cite{parm94}--\cite{parm99}, CHARMM \cite{charmm}, OPLS \cite{opls1,opls2}, 
GROMOS \cite{gromos}, and ECEPP \cite{ECEPP}.
These force fields have been parameterized to fit experimental
data of small molecules
and, for some terms, quantum chemistry calculations.
In fact, it is rather difficult to test the validity of these
force-field parameters, because simulations of biomoleuclar
systems will be hampered by the multiple-minima problem; the simulations
will get trapped in states of energy local minima and will yield
unreliable results.
Hence, it is essential to employ powerful simulation algorithms 
such as generalized-ensemble algorithms (for a recent review, see
Ref.~\cite{GEA}) in order to have accurate comparisons of 
force fields.   For instance, we have recently carried out
detailed comparisons of
three versions (parm94 \cite{parm94}, parm96 \cite{parm96},
and parm99 \cite{parm99}) of AMBER, CHARMM \cite{charmm},
OPLS-AA/L \cite{opls2}, and GROMOS \cite{gromos}
by generalized-ensemble simulations of small peptides \cite{YSO}.
   
If any problems are found in the force fields, we have to
improve the force-field parameters by some means.
In this Letter, we propose a novel method for optimization of
force-field parameters.
Our method utilizes the protein structures in the Protein Data Bank
(PDB) \cite{PDB},
in which the number of entries is rapidly increasing.
Actually, there already exist many works 
of similar knowledge-based methods of force-field development
\cite{other1}--\cite{other9}.
Our method is different from most of previous works in two points: we deal with all-atom
force fields such as AMBER, and we use only the PDB data without
introducing so-called decoys (or, misfolded structures).

In section 2 the details of the formulation of our method are described.
In section 3 an example of the application of our method is presented.
Conclusions and future prospects are presented in section 4.

\section{Methods}
\label{method}
The existing all-atom force fields for protein systems such as AMBER
and CHARMM use essentially the same functional forms for the conformational
potential energy $E_{\rm conf}$ except for
minor differences.  $E_{\rm conf}$ can be written as, for instance,
% Here, we use the following functions for
%the potential energyThere are 
%We calculate the conformational energy by the force field 
%equation of AMBER (\ref{ene_conf}).

\begin{equation}
E_{\rm conf} = E_{\rm BL} + E_{\rm BA} + E_{\rm torsion} + E_{\rm nonbond}~,
\label{ene_conf}
\end{equation}

\begin{equation}
E_{\rm BL} = \sum_{{\rm bond~length}~\ell} K_{\ell} (\ell - \ell_{\rm eq})^2~,
\label{ene_bond}
\end{equation}
\begin{equation}
E_{\rm BA} = \sum_{{\rm bond~angle}~\theta} K_{\theta} ( \theta - \theta_{\rm eq})^2~,
\label{ene_angle}
\end{equation}
\begin{equation}
E_{\rm torsion} = \sum_{{\rm dihedral~angle}~\Phi} \sum_n \frac{V_n}{2} [ 1 + \cos (n \Phi - \gamma_n) ]~,
\label{ene_torsion}
\end{equation}
\begin{equation}
E_{\rm nonbond} = \sum_{i<j} \left[ \frac{A_{ij}}{r_{ij}^{12}} - \frac{B_{ij}}{r_{ij}^6} 
+ \frac{q_i q_j}{\epsilon r_{ij}} \right] .
\label{ene_nonbond}
\end{equation}

Here, $E_{\rm BL}$, $E_{\rm BA}$, and $E_{\rm torsion}$
represent the bond-stretching term, the 
bond-bending term, and the torsion-energy term, respectively.
The bond-stretching and bond-bending energies are given by harmonic terms 
with the force constants $K_{\ell}$ and $K_{\theta}$, and the equilibrium positions, 
$\ell_{\rm eq}$ and $\theta_{\rm eq}$.
The torsion energy is, on the other hand, described by the Fourier 
series in Eq.~(\ref{ene_torsion}), 
where the sum is taken over all dihedral angles $\Phi$,
$n$ is the number of waves, $\gamma_n$ is the phase, and $V_n$ is 
the Fourier coefficient.
The nonbonded energy in Eq.~(\ref{ene_nonbond}) is represented by 
the  Lennard-Jones and Coulomb terms between pairs of atoms, $i$ and 
$j$, separated by the distance $r_{ij}$.
The parameters $A_{ij}$ and $B_{ij}$ in Eq. (\ref{ene_nonbond}) are the coefficients
for the Lennard-Jones term, 
$q_i$ is the partial charge of the $i$-th atom in the electrostatic term,
and $\epsilon$ is the dielectric constant, where we set $\epsilon = 1$.
Hence, we have five classes of force-field parameters, namely,
those in the bond-stretching term ($K_{\ell}$ and $\ell_{\rm eq}$), those in 
the bond-bending term
($K_{\theta}$ and $\theta_{\rm eq}$), those in the torsion term
($V_n$ and $\gamma_n$), those in the Lennard-Jones term
($A_{ij}$ and $B_{ij}$), and those in the electrostatic
term ($q_i$).

We now describe our new method for optimizing these
force-field parameters.
We first select $N$ molecules in PDB.  We try to choose proteins
from different folds
(such as all $\alpha$-helix, all $\beta$-sheet, $\alpha$/$\beta$, etc.)
and different homology classes
as much as possible.
If the force-field parameters are of ideal values, we expect that
all the chosen native structures are stable without any force acting on
each atom in the molecules.  Hence, we expect
\begin{equation}
F = 0~,
\label{F0}
\end{equation}
where
\begin{equation}
%F = \sum_{m=1}^N \frac{1}{N_m} \sum_{i_m = 1}^{N_m} \left| \overrightarrow{f}_{i_m} \right| ^2~,
F = \sum_{m=1}^N \frac{1}{N_m} \sum_{i_m = 1}^{N_m} \left| \vec{f}_{i_m} \right| ^2~,
\label{F}
\end{equation}
and
\begin{equation}
%\overrightarrow{f}_{i_m} = - \frac{\partial E_{\rm tot}}{\partial \overrightarrow{x}_{i_m}}~. 
\vec{f}_{i_m} = - \frac{\partial E_{\rm tot}}{\partial \vec{x}_{i_m}}~. 
\label{F1}
\end{equation}
Here, $N_m$ is the total number of atoms in molecule $m$, $E_{\rm tot}$ is
the total potential energy, and $\vec{f}_{i}$ is 
the force acting on atom $i$.
In reality, $F \ne 0$, and 
because $F \ge 0$, we expect that we can optimize the force-field parameters
by minimizing $F$ with respect to these parameters.
In practice, we perform a simulation in the force-field parameter space for this
minimization.

Proteins are usually in aqueous solution, and hence we also have to incorporate
some kind of solvent effects.  Because the more the total number of proteins
($N$) is, the better the force-field parameter optimizations are expected to be, we 
want to minimize our efforts in the calculations of the solvent effects.
Here, we employ the generalized-Born/surface area (GB/SA) terms
for the solvent contributions \cite{gb1,gb2}.  
Hence, we use in Eq.~(\ref{F1})
\begin{equation}
E_{\rm tot} = E_{\rm conf} + E_{\rm solv}~,
\label{ene_tot}
\end{equation}
where
\begin{equation}
E_{\rm solv} = E_{\rm GB} + E_{\rm SA}~,
\label{ene_gbsa1}
\end{equation}
\begin{equation}
E_{\rm GB} = - 166 \left( 1 - \frac{1}{\epsilon_{s}} \right) \sum_{i,j} \frac{q_i q_j}
%E_{\rm GB} = - 166 \left( 1 - \frac{1}{\epsilon_{s}} \right) \sum_{i=1}^n \sum_{j=1}^n \frac{q_i q_j}
%{(r_{ij}^2 + \alpha_{ij}^2 e^{-D_{ij}})^{0.5}}~,
{\sqrt{r_{ij}^2 + \alpha_{ij}^2 e^{-D_{ij}}}}~,
\label{ene_gbsa3}
\end{equation}
\begin{equation}
E_{\rm SA} = \sum_k \sigma_k A_k~.
\label{ene_gbsa2}
\end{equation}
Namely, in the GB/SA model, the total solvation free energy 
in Eq.~(\ref{ene_gbsa1}) is given by the sum 
of a solute-solvent electrostatic 
polarization term,
a solvent-solvent 
cavity term, and a solute-solvent van der Waals term.
A solute-solvent electrostatic polarization term can be calculated by the generalized Born 
equation (\ref{ene_gbsa3}), where
$\alpha_{ij} = \sqrt{\alpha_i \alpha_j}$,
$\alpha_i$ is the so-called Born radius of atom $i$,
$D_{ij} = r_{ij}^2/(2\alpha_{ij})^2$, and
$\epsilon_s$ is the dielectric constant of bulk water (we take $\epsilon_s = 78.3$).
%and the double 
%sum runs over all pairs of atoms ($i$ and $j$). $\alpha_i$ is the so-called Born radius of atom $i$.
%$D_{ij}$ is the squared ratio of the $i$ and $j$th atom pair separation to their mean Born diameters.
%The $\epsilon_s$ is the dielectric constant of bulk water, and we take $\epsilon_s = 78.3$.
A solvent-solvent cavity term and a solute-solvent van der Waals term can 
be approximated by the term that is proportional to the 
solvent accessible surface area in Eq.~(\ref{ene_gbsa2}).
Here,
$A_k$ is the total solvent-accessible surface area of atoms of type $k$ and $\sigma_k$ is 
an empirically determined proportionality constant
%, and we take $\sigma_k = 4.9$ cal/mol 
\cite{gb1,gb2}.

\begin{figure}
\begin{center}
\includegraphics[width=16cm,keepaspectratio]{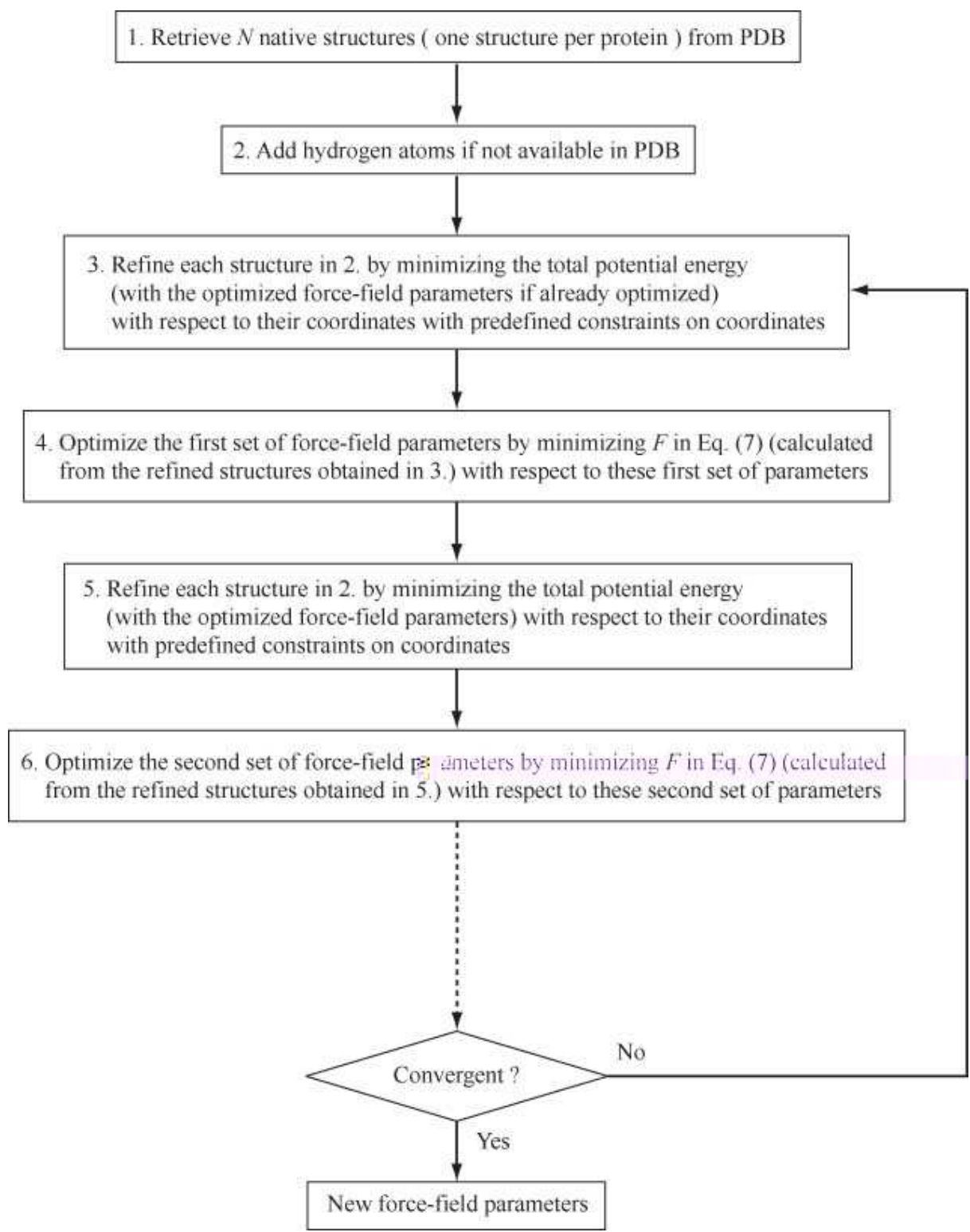}
%\caption{The flowchart of the optimization of force-field parameters.}
\caption{The flowchart of our method for the optimization of force-field parameters.} 
\label{fig_1}
\end{center}
\end{figure}

The flowchart of our method for the optimization of force-field parameters 
is shown in Fig.~\ref{fig_1}.

%\begin{itemize}
%\item[1.] Native structure of protein (PDB) : \\
In Step 1 we try to obtain as many structures as possible from PDB.
The number is limited by the computer power that we have available in our laboratory.
We want to choose proteins with different sizes (numbers of amino acids), 
different folds, and different homology classes as much as possible.
We also want to use only those with high experimental resolutions.
Note that only atomic coordinates of proteins are extracted from PDB
(and coordinates from other molecules such as crystal water are neglected).
%We have to be careful to 
%All proteins are obteined from PDB (Protein Data Bank) determined by X-ray diffractions.
%PDB data should be chosen with high resolutions and various scale proteins.

% PDB files for the optimization are the number of 100, 200 or less residues, and 
% the resolution of all data is 1.8 or less $\AA$. 

%\item[2.] Add hydrogen atom if not available : \\
If we use data from X-ray experiments, hydrogen atoms are missing, and thus in
Step 2 we have to add hydrogen coordinates.
Many protein simulation software packages provide with routines that add hydrogen atoms to
the PDB coordinates, and one can use one of such routines.  
%Here, we use that in
%Hydrogen atoms do not exist in PDB data determined by X-ray diffraction, and they are made 
%artifically by 
%TINKER \cite{tinker1}--\cite{tinker7}.

%\item[3.] Refinement of input structure : \\
We now have $N$ protein coordinates ready, but usually such ``raw data''
result in very high total potential energy and strong forces will be
acting on some of the atoms in the molecules. 
This is because the hydrogen coordinates we added as above are not based on
experimental results and have rather large uncertainties.
The coordinates of heavy atoms from PDB also have experimental uncertainties.
We take the position that we leave the coordinates of heavy atoms as they are
in PDB as much as possible, and adjust the hydrogen coordinates to
adjust this mismatch.
This is why we want to include as many PDB data as possible with high
experimental resolution (so that the effects of experimental errors
in PDB may be minimal).
We thus minimize the total potential energy 
$E_{\rm tot} = E_{\rm conf} + E_{\rm solv} + E_{\rm constr}$
with respect to the coordinates for each protein conformation,
where $E_{\rm constr}$ is the constraint energy term that is imposed on
the heavy atoms in PDB (it is referred to as the ``predefined constraints''
in Steps 3 and 5 in Fig.~\ref{fig_1}):
%Poteintial energies of all PDB data are minimized  with the restriction except hydrogen atoms.
%Their coordinates are the random position, therefore the energy and its derivatives are 
%very diffrent from native structures.
%Derivatives of the original PDB data and the minimization data are shown in Fig. ().
%RMSD (Root Mean Square Distance) of this two structures of the molecule (2LIS) is 0.027 $\AA$, 
%however the scale of their deviratives is the average of 8.209 $kcal/mol \AA$.
\begin{equation}
%E_{\rm constr} = \sum_{\rm heavy~atom} K_x (\overrightarrow{x} - \overrightarrow{x_0})^2~.
E_{\rm constr} = \sum_{\rm heavy~atom} K_x (\vec{x} - \vec{x}_0)^2~.
\label{ene_constr}
\end{equation}
Here,
%$E_{\rm constr}$ is the constrain energy of the restriction except hydrogen atoms.
$K_x$ is the force constants of the restriction (in the present work, we use the value of 
100 kcal/mol/\AA$^2$), and
$\vec{x}_0$ are the original coordinate vectors of heavy atoms in PDB.
Because we are searching for the nearest local-minimum states,
usual minimization routines such as conjugate-gradient method and 
Newton-Raphson method can be employed here.
As one can see from Eq.~(\ref{ene_constr}), the coordinates of hydrogen
atoms will be mainly adjusted, but unnatural heavy-atom coordinates
will also be modified.
We perform this minimization for all $N$ protein structures separately, and
obtain $N$ refined structures.

%\item[4.] Optimization of first set of parameters : \\
Given $N$ set of ``ideal'' reference coordinates in Step 3,
we now optimize the first set of force-field parameters in Step 4.
In Eq.~(\ref{ene_conf}) we have five classes of force-field parameters as mentioned
above.
Namely, the force-field parameters are
those in the bond-stretching term ($K_{\ell}$ and $\ell_{\rm eq}$), those in 
the bond-bending term
($K_{\theta}$ and $\theta_{\rm eq}$), those in the torsion term
($V_n$ and $\gamma_n$), those in the Lennard-Jones term
($A_{ij}$ and $B_{ij}$), and those in the electrostatic
term ($q_i$).
Because they are of very different nature, we believe that it is better to
optimize these classes of force-field parameters separately (as in
Steps 4, 6, and so on in Fig.~\ref{fig_1}).
For each set of force-field parameters, the optimization is carried out
by minimizing $F$ in Eq.~(\ref{F}) with respect to these parameters.
Here, $E_{\rm tot}$ in Eq.~(\ref{F1}) is given by Eq.~(\ref{ene_tot})
For this purpose usual minimization routines such as conjugate-gradient
method are not adequate, because we need a global optimization.
One can employ more powerful methods such as simulated annealing \cite{SA}
and generalized-ensemble algorithms \cite{GEA}.
We perform this minimization simulation
in the above parameter space to obtain the parameter
values that give the global minimum of $F$.

These processes are repeated until the optimized force-field parameters
converge.

We can, in principle, optimize all the force-field parameters following the flowchart
in Fig.~\ref{fig_1}.
In the example given in the next section, however, we just optimize two classes
of the force-field parameters; namely, the partial charges ($q_i$) and the
backbone torsion energy parameters ($V_1$, $V_2$, and $V_3$).
Here, we fix the phases ($\gamma_n$) and the main-chain torsion energy term is
written as (see Eq.~(\ref{ene_torsion}))
\begin{equation}
E_{\Phi=\phi,\psi} = \frac{V_1}{2} \left[ 1+\cos (\Phi) \right] 
+ \frac{V_2}{2} \left[ 1+\cos (2\Phi - \pi) \right] + \frac{V_3}{2} \left[ 1+\cos (3\Phi) \right]~,
\label{ene_main}
\end{equation}
for each backbone dihedral angle $\phi$ and $\psi$.
As for the partial charge optimization, 
we impose a condition that the total charge of each amino acid remains constant,
which is the usual assumption adopted by force fields of 
Eqs.~(\ref{ene_bond})--(\ref{ene_nonbond})
based on classical mechanics.

We believe that these two classes of parameters have the most uncertainty among 
all the force-field parameters. 
This is because partial charges are usually obtained by quantum chemistry
calculations of an isolated amino acid in vacuum separately, which is a very
different condition from that in amino acids of proteins in aqueous solution
and because torsion-energy term is the most problematic (for instance, parm94,
parm96, and parm99 versions of AMBER differ mainly in torsion-energy parameters).
%We use charge parameters for the first parameters, and use backbone dihedral parameters for 
%the second parameters.
%The optimization of charge parameters of Eq.(\ref{ene_gbsa3}) and 
%Eq.(\ref{ene_nonbond}) by MC simulated annealing method is performed.
%In addition, this simulatin are performed with the restriction that the total charge of 
%each amino acid is a constant value.
%The optimization of backbone dihedral parameters ($\phi$ and $\psi$) is performed.
%We use three variables, $V_1$, $V_2$, and $V_3$ (Eq.(\ref{ene_main})) for
%each backbone dihedral angle $\phi$ and $\psi$.
%\begin{equation}
%E_{\Phi=\phi,\psi} = \frac{V_1}{2} \left[ 1+\cos (\Phi) \right] 
%+ \frac{V_2}{2} \left[ 1+\cos (2\Phi - \pi) \right] + \frac{V_3}{2} \left[ 1+\cos (3\Phi) \right]~.
%\label{ene_main}
%\end{equation}

\section{Results and discussion}
\label{results}

We now present an example of the application of our force-field optimization 
scheme presented in the previous section.
The force field that we optimized is the AMBER parm96 version \cite{parm96}.
We have optimized two sets of parameters.  The first set is the partial
charge parameters ($q_i$ in Eq.~(\ref{ene_nonbond})) and there are 602 such
parameters altogether in AMBER.
The second set is the backbone torsion-energy parameters
($V_1$, $V_2$, and $V_3$ in Eq.~(\ref{ene_main})) and there are six such parameters
(three each for $\phi$ and $\psi$).
We used the program package TINKER \cite{tinker} for all the calculations in
the present work.

In Step 1 of the flowchart of Fig.~\ref{fig_1},
we chose 100 PDB files ($N=100$) with resolution 1.8 \AA~ or better and with
less than 200 residues 
(the average number of resiudes is 120.4) from PISCES \cite{pisces}.
%The database contains $100$ proteins listed below (in PDB code names (chain names)): 
Their PDB code names are
% database contains $100$ proteins listed below (in PDB code names): 
%2LIS(A) 1EP0(A) 
%1TIF 1EB6(A) 1C1L(A) 1CCW(A) 2PTH 1I6W(A) 1DBF(A) 1KPF 1LRI(A) 1AAP(A) 1C75(A) 1CC8(A) 
%1FK5(A) 1KQR(A) 1K1E(A) 1CZP(A) 1GP0(A) 1KOI(A) 1IQZ(A) 3EBX 1I40(A) 1EJG(A) 1AMM 1I07(A) 
%1GK8(I) 1GVP 1M4I(A) 1EYV(A) 1E29(A) 1I2T(A) 1VCC 1FM0(D) 1EXR(A) 1GUT(A) 1H4X(A) 1GBS 
%1B0B 119L 1IFC 1DLW(A) 1EAJ(A) 1GGZ(A) 1JR8(A) 1RB9 1VAP(A) 1JZG(A) 1M55(A) 1EN2(A) 
%1C9O(A) 2ERL 1EMV(A) 1F41(A) 1EW6(A) 2TNF(A) 1IFR(A) 1JSE 1KAF(A) 1HZT(A) 1HQK(A) 1FXL(A) 
%1BKR(A) 1ID0(A) 1LQV(B) 1G2R(A) 1KR7(A) 1QTN(A) 1D4O(A) 1EAZ(A) 2CY3 1UGI(A) 1IJV(A) 
%3VUB 1BZP(A) 1JYR(A) 1DZK(A) 1QFT(A) 1UTG 2CPG(A) 1I6W(A) 1C7K(A) 1I8O(A) 1LO7(A) 1LNI(A) 
%1EQO(A) 1NDD(A) 1HD2(A) 3PYP 1FD3(A) 1DK8(A) 1WHI 1FAZ(A) 4FGF 2MHR 1JB3(A) 2MCM 1IGD 
%1C5E(A) 1JIG(A).
2LIS, 1EP0, 
1TIF, 1EB6, 1C1L, 1CCW, 2PTH, 1I6W, 1DBF, 1KPF, 1LRI, 1AAP, 1C75, 1CC8, 
1FK5, 1KQR, 1K1E, 1CZP, 1GP0, 1KOI, 1IQZ, 3EBX, 1I40, 1EJG, 1AMM, 1I07, 
1GK8, 1GVP, 1M4I, 1EYV, 1E29, 1I2T, 1VCC, 1FM0, 1EXR, 1GUT, 1H4X, 1GBS, 
1B0B, 119L, 1IFC, 1DLW, 1EAJ, 1GGZ, 1JR8, 1RB9, 1VAP, 1JZG, 1M55, 1EN2, 
1C9O, 2ERL, 1EMV, 1F41, 1EW6, 2TNF, 1IFR, 1JSE, 1KAF, 1HZT, 1HQK, 1FXL, 
1BKR, 1ID0, 1LQV, 1G2R, 1KR7, 1QTN, 1D4O, 1EAZ, 2CY3, 1UGI, 1IJV, 
3VUB, 1BZP, 1JYR, 1DZK, 1QFT, 1UTG, 2CPG, 1I6W, 1C7K, 1I8O, 1LO7, 1LNI, 
1EQO, 1NDD, 1HD2, 3PYP, 1FD3, 1DK8, 1WHI, 1FAZ, 4FGF, 2MHR, 1JB3, 2MCM, 
1IGD, 1C5E, and 1JIG.

\begin{figure}
\begin{center}
\includegraphics[width=17cm,keepaspectratio]{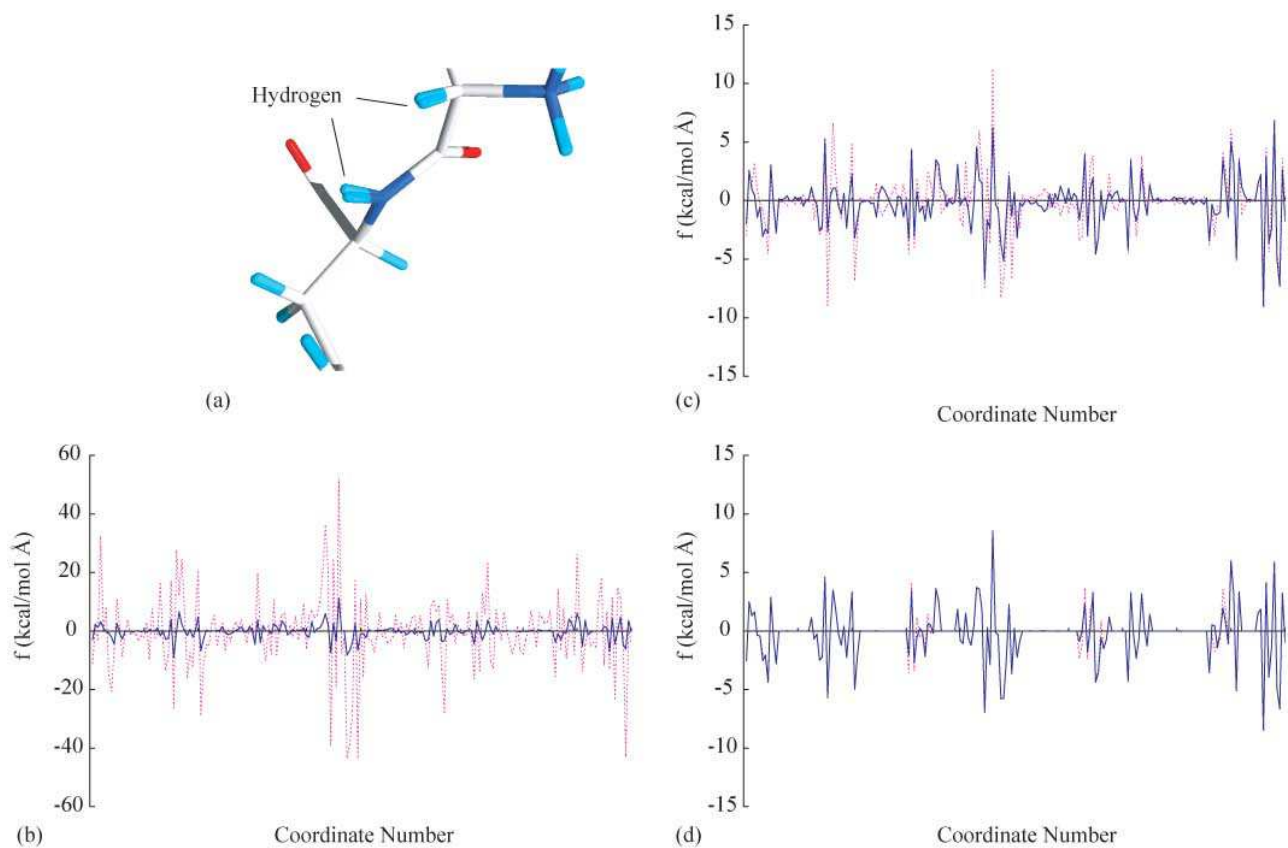}
\caption{(a) Superposition of a part of the original PDB structure (PDB code: 2LIS)
with hydrogen atoms added and the corresponding part in the refined
structure.  These structures were obtained in Steps 2 and 3 of the flowchart in
Fig.~\ref{fig_1}.
(b) Components of the force acting on the atoms in the original structure of 2LIS
(red dotted curve) and its refined structure (blue solid curve).  The structures
are the same as those in (a).  The force
field is the original AMBER parm96.  
(c) Components of the force acting on the atoms in 
the refined structure of 2LIS.  The force field is the original AMBER parm96
(red dotted curve) and AMBER parm96 with the partial-charge parameters optimized
(blue solid curve).  
(d) Components of the force acting on the atoms in 
the refined structure of 2LIS.  The force field is the original AMBER parm96
with the partial-charge parameters optimized
(red dotted curve) and AMBER parm96 with the partial-charge and torsion-energy 
parameters optimized 
(blue solid curve).  In (b), (c), and (d), only 200 components are shown.}
\label{fig_2}
\end{center}
\end{figure}

%In order to improve the accuracy of the force-field parameters, we have to prepare PDB files 
%reflecting the rate of the amino acid sequence of a nature.
%However, the rate of the sequence is not considered to these PDB files.
%For their PDB data, hydragen atoms were added artifically, and so the minimization was performed 
%with the restriction except hydrogen atoms using the original AMBER parm96 force field parameters.
The minimization for the coordinate refinement in Step 3 of the flowchart
was done with the constraint of Eq.~(\ref{ene_constr}).
%by the quasi Newton method \cite{new1}--\cite{new3}, which is implemented
%in TINKER, 
%We used the convergence criterion that
%RMSG (Root Mean Square Gradient) of energy 
%is $0.01$ kcal/mol \AA~ or less. 
In Fig. \ref{fig_2}(a), a part of the original PDB structure of one of the molecules (2LIS) 
and the corresponding part in the refined structure 
are superposed.
As expected, we see good coincidence of heavy-atom coordinates and small deviations
in the hydrogen coordinates.
%Derivatives of the original PDB structure of 2LIS and the minimization structure of that are shown in Fig. 
%\ref{fig_2}b.
RMSD (Root Mean Square Distance) (for heavy atoms)
of these two conformations of the entire molecule (2LIS) 
is indeed small: 0.027 \AA~ (the average of all 100 proteins is 0.03 \AA). 
  
As also expected, the structure of the raw PDB data (that obtained in Step 2 of
the flowchart) is not stable in the sense that the force acting
on some atoms in the molecule will be large, while the refined structure
(that obtained in Step 3 of the flowchart) will not have abnormally large
forces acting.  This is illustrated in Fig.~\ref{fig_2}(b).
The average absolute value of 
the components of the force before minimization and after minimization
is 9.0 kcal/mol/\AA~ and 1.5 kcal/mol/\AA, respectively.
%By this procedure, their PDB data become more near native structure.

\begin{figure}
\begin{center}
\includegraphics[width=7cm,keepaspectratio]{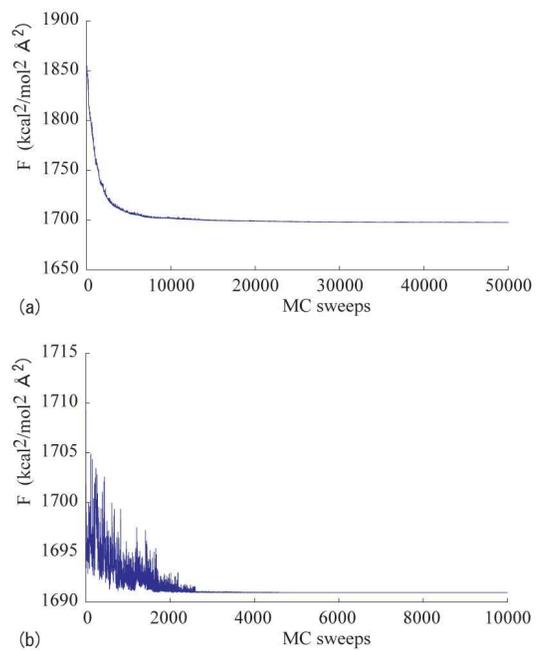}
\caption{Time series of MC simulated annealing simulations in force-field parameter
space for partial-charge (a) and torsion-energy (b) parameters.} 
\label{fig_3}
\end{center}
\end{figure}

In Step 4 of the flowchart,
we performed the optimization of partial-charge parameters by 
MC simulated annealing.
Namely, we minimized $F$ in Eq.~(\ref{F}) by 
MC simulated annealing simulations of these parameters
(the parameters are updated and the updates are accepted or rejected
according to the Metropolis criterion).
%The number of the variable of the charge parameter is $602$.
For this we introduced an effective ``temperature'' for the parameter space.
Each simulation run consisted of 50,000 MC sweeps with the temperature
decreased exponentially from 200 to 0.01.
The simulation was repeated 10 times with different initial random numbers.
The time series of $F$ from one of the simulations is
shown in Fig. \ref{fig_3}(a).
We see that $F$ decreases quickly in the beginning until about 5,000
MC sweeps and then it decreases very slowly; the total number of
MC sweeps (50,000) seems sufficient.
The optimized partial charges are taken from those that resulted in
the lowest $F$ value.

Each term contributing to $F$ (i.e., each component of the force acting on the
atoms) before and after the optimization is compared in Fig.~\ref{fig_2}(c).
%Change of derivatives with optimized charge parameters is shown in Fig. \ref{fig_2}c.
We see that many terms decreased in magnitude as a whole.

In Tables \ref{table-ala}, \ref{table-glu}, and \ref{table-tyr},
three examples (alanine, glutamic acid, and tyrosine) of the obtained partial
charges together with the original AMBER values are listed.
The magnitudes of the charges seem to decrease a little in general, although
there are exceptions (see, for instance, CG and CE of tyrosine).
%Three tables show that the values of the partial charges have not changed a lot.
Although the sign of the partial charges remains the same for those with large magnitude,
charges with small magnitude sometimes change their sign
(see, for example, CG of glutamic acid and CA, CB, and HB 
of tyrosine).

In Step 5 of the flowchart, 
the original coordinates obtained in Step 2 were again minimized 
with the constraint of Eq.~(\ref{ene_constr}),
but this time the optimized parameters were used.
%Using optimized parameters, the minimization was performed with the restriction except hydrogen atoms.
The average RMSD of 100 proteins is 0.03 \AA, and the coordinates of heavy atoms
have little changed.

In Step 6 of the flowchart, 
we carried out the optimization of the six torsion-energy parameters 
($V_1$, $V_2$, and $V_3$ in Eq.~(\ref{ene_main}) for both $\phi$ and $\psi$) 
by minimizing $F$ in Eq.~(\ref{F}) with
MC simulated annealing simulations in this parameter space.
%The number of the variable of the dihedral parameter is $6$.
Each simulation run consisted of 10,000 MC sweeps with the temperature 
decreasing from 1,000 to 1.0.
The simulation was repeated several times with different random numbers, but
the six optimized parameters all agreed with one another.
%however all the values of dihedral parameters of results 
%were the same.
The time series of $F$ from one of the simulations is
shown in Fig. \ref{fig_3}(b),
%One of results of simulations is shown in Fig. \ref{fig_3}b, 
and the obtained torsion-energy parameters 
are listed in Table~\ref{table-torsion}.
Each term contributing to $F$ 
before and after the optimization is compared in Fig.~\ref{fig_2}(d).
The term correspoding to the backbone atoms that are relevant to the torsion-energy
parameters decreased, as expected.
Non-zero contributions are presumably to be reduced by optimization of
other force-field parameters.

In the present example, we stopped our process in Step 6 of the flowchart
and did not iterate the optimizations.

We tested the effectiveness of the new force field by applying it to the
folding simulations of two peptides, C-peptide of ribonuclease A 
and the C-terminal fragment of the B1 domain of
streptococcal protein G,
which is sometimes referred to as G-peptide \cite{gpep3}. 
The C-peptide has 13 residues and its amino-acid sequence is
Lys-Glu-Thr-Ala-Ala-Ala-Lys-Phe-Glu-Arg-Gln-His-Met.
This peptide has been extensively studied by experiments and is known
to form an $\alpha$-helix structure \cite{buzz1,buzz2}.
Because the charges at peptide termini are known to affect
helix stability \cite{buzz1,buzz2}, we blocked the termini by
a neutral COCH$_3$- group and a neutral -NH$_2$ group.
%Total sequence is Ace-Lys-Glu-Thr-Ala-Ala-Ala-Lys-Phe-Glu-Arg-Gln-His-Met-NH2.
%This peptide is known as $\alpha$-helix peptide experimentally.
The G-peptide has 16 residues and its amino-acid sequence is
Gly-Glu-Trp-Thr-Tyr-Asp-Asp-Ala-Thr-Lys-Thr-Phe-Thr-Val-Thr-Glu.
The termini were kept as the usual zwitter ionic states, following the
experimental conditions \cite{gpep1,gpep2,gpep3}.
This peptide is known to form a $\beta$-hairpin structure by experiments
\cite{gpep1,gpep2,gpep3}.
      
Simulated annealing MD simulations of 1 ns were performed for both
peptides from extended initial conformations.
The unit time step was set to 1.0 fs (hence, each run consists of
1,000,000 MD steps).
The temperature during MD simulations was controlled by 
Berendsen's method \cite{berendsen}.
For each simulation the temperature was decreased from 2,000 K to 200 K.
As for solvent effects, in order to carry out quantitative test,
we want to include accurate contributions such as explicit water
molecules.  Here, for simplicity, however, we used the GB/SA model in 
Eqs.~(\ref{ene_gbsa1})--(\ref{ene_gbsa2}) was used.
For both peptides, these folding simulations were repeated 16 times 
with different initial conformations.
As for force fields, both the original AMBER parm96 force field and
the optimized force field were used. 
%For the judgment of the second structure, the method of DSSP\cite{DSSP} was used.

\begin{figure}
\begin{center}
\includegraphics[width=17cm,keepaspectratio]{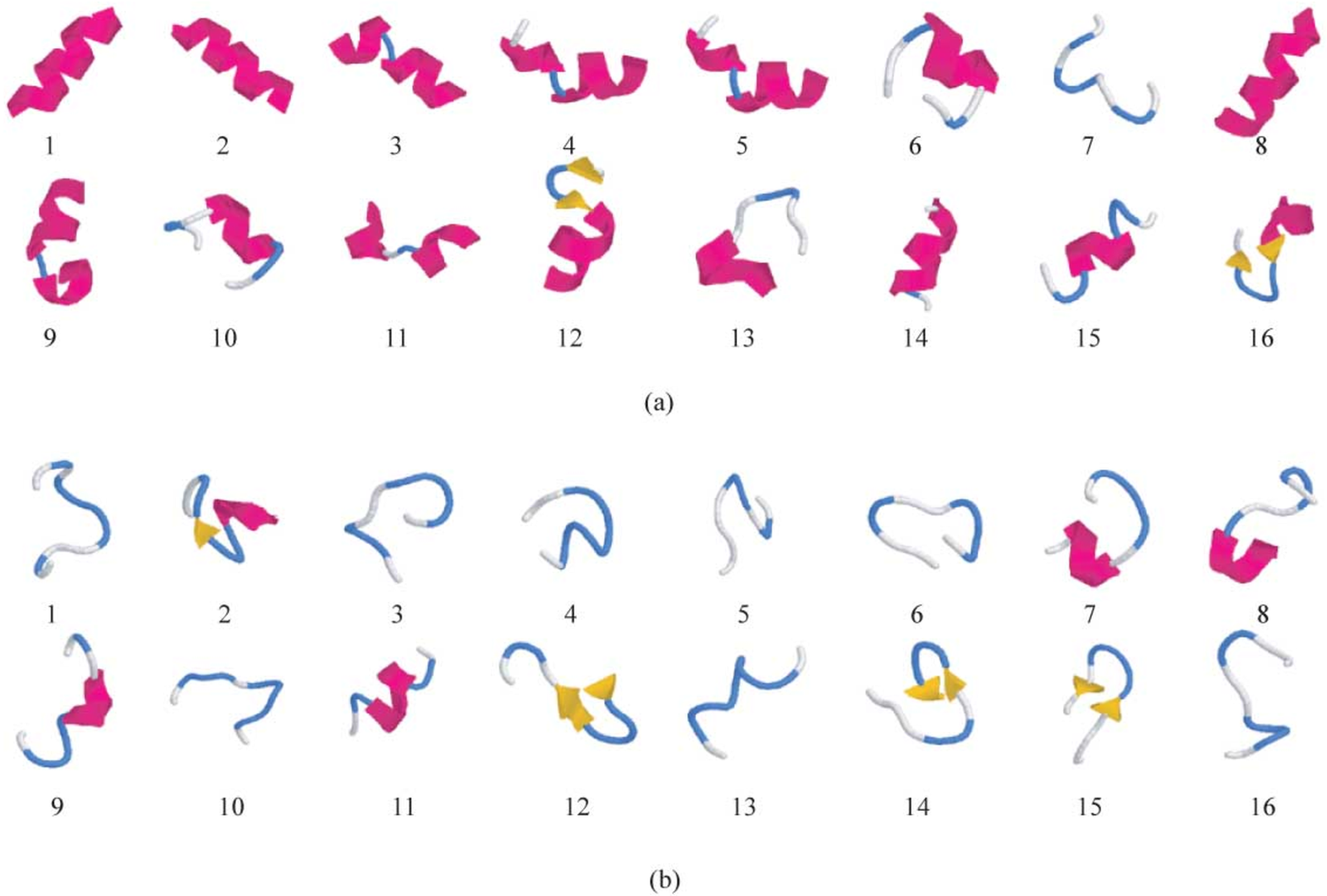}
\caption{The lowest-energy conformations of C-peptide obtained by each of the
simulated annealing MD simulations using the original AMBER parm96 (a) and 
the optimized force field (b).  The conformations are ordered in the increasing
order of energy.  The figures were created with RasMol \cite{rasmol}.}
\label{fig_4}
\end{center}
\end{figure}

In Fig.~\ref{fig_4} the lowest-energy conformations of C-peptide obtained 
from each of the folding simulations 
are shown.  They are numbered in increasing order of energy.
%from the low structure (a: original AMBER parm96, b: optimized force field).
Almost all conformations are $\alpha$-helix structure in the case of 
the original AMBER parm96 (see Fig.~\ref{fig_4}(a)), although 
a short $\beta$-hairpin structure is
also found in two conformations (Nos. 12 and 16).
%the hydrogen-bonded of $\beta$-hairpin appeared in No.12 and 16.
%In the case of the original AMBER parm96, the structure where the value of 
RMSD (only C$_{\alpha}$ atoms are taken into account) from the corresponding
part of the native
X-ray structure of the entire ribonulcease A was
the lowest for conformations Nos. 3 (2.47 \AA), 14 (2.59 \AA), and 15
(2.81 \AA).
%The average total potential energy of $16$ structures is $-569.78$ kcal/mol, 
%the average electrostatic energy is $-332.22$ kcal/mol, and the average torsional energy 
%is $48.08$ kcal/mol.
For the optimized force field (see Fig.~\ref{fig_4}(b)), 
five out of 16 conformations (Nos. 2, 7, 8, 9, and 11)
are $\alpha$-helix structure, and $\beta$-hairpin structure also
appeared in Nos. 12, 14, and 15.
Though both $\alpha$-helix and $\beta$-hairpin structures appear
with the optimized force field, $\alpha$-helix is more favored than
$\beta$-hairpin in the sense that the energy is lower for conformations
with $\alpha$-helix than those with $\beta$-hairpin.
%The structure where the value of RMSD is the lowest is No.$13$.
RMSD was the lowest for conformations Nos. 13 (3.17 \AA), 16 (3.50 \AA), and 11
(3.51 \AA).
%The average total energy of $16$ structures is $-648.29$ kcal/mol, 
%the average electrostatic energy is 
%$-430.69$ kcal/mol, and the average torsional energy is $65.69$ kcal/mol.
Although the original AMBER parm96 gave more conformations closer to
the native structure (smaller RMSD) than the optimized force field, 
it seems to favor $\alpha$-helix too much.  The experiments imply
only 30 \% $\alpha$-helix formations around $0^{\circ}$ C \cite{buzz1},
which is more consistent with the results for the optimized force field.

\begin{figure}
\begin{center}
\includegraphics[width=17cm,keepaspectratio]{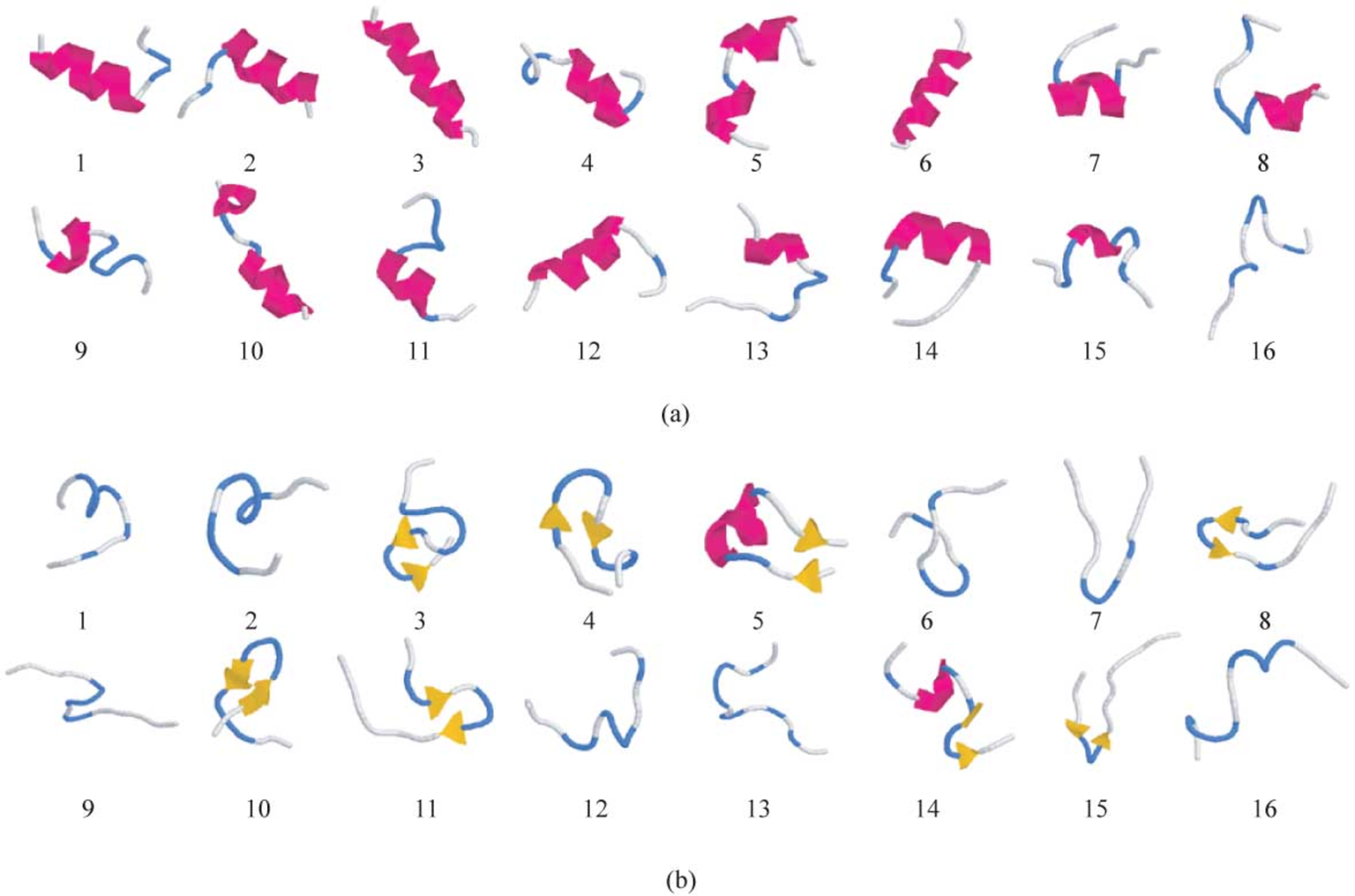}
\caption{The lowest-energy conformations of G-peptide obtained by each of the
simulated annealing MD simulations using the original AMBER parm96 (a) and 
the optimized force field (b).  The conformations are ordered in the increasing
order of energy.  The figures were created with RasMol \cite{rasmol}.}
\label{fig_5}
\end{center}
\end{figure}

In Fig.~\ref{fig_5} the lowest-energy conformations of G-peptide obtained 
from the folding simulations are shown.  
%Experimentally, G-peptide is $\beta$-hairpin structure, whereas in the case of 
The results for the original AMBER parm96 (see Fig.~\ref{fig_5}(a)), 
are again
almost all $\alpha$-helical.  This is in contradiction with the experimental
results \cite{gpep1,gpep2,gpep3} which imply the formation of a $\beta$-hairpin.
RMSD from the corresponding
part of the native
X-ray structure of the entire protein G was
the lowest for conformations Nos. 14 (4.52 \AA), 13 (5.30 \AA), and 8
(5.53 \AA).
%In the case of the original AMBER parm96, the structure where 
%the value of RMSD is the lowest is No.$14$.
%The average total potential energy of $16$ structures is $-659.80$ kcal/mol, 
%the average electrostatic energy is $-239.84$ kcal/mol, and the average torsional energy 
%is $41.66$ kcal/mol.
For the optimized force field (see Fig.~\ref{fig_5}(b)), 
%In that of the optimized force field, structures more than a half 
as many as eight out of 16 conformations (Nos. 3, 4, 5, 8, 10, 11, 14, and 15) 
yielded $\beta$-hairpin structure, and $\alpha$-helix structure
also appeared in Nos. 5 and 14.
RMSD was the lowest for conformations Nos. 7 (3.74 \AA), 4 (3.84 \AA), and 11
(4.00 \AA).
Hence, for the case of G-peptide, the improvement of the force field by
our optimization procedure is great (both in secondary structure content
and in RMSD).
%The structure where the value of RMSD is the lowest is No.$7$.
%The average total energy of $16$ structures is $-785.56$ kcal/mol, 
%the average electrostatic energy is 
%$-373.99$ kcal/mol, and the average torsional energy is $61.60$ kcal/mol.

%Compared with the original AMBER parm96, it turned out that the optimized force field becomes 
%the force field which is easy to take $\beta$-hairpin structure rather than $\alpha$-helix structure. 
%Although, as for the optimized force field, the value of electrostatic energy is decreasing 
%compared with the original AMBER parm96, the value of the torsional energy is increasing.

\section{Conclusions}
\label{conclusions}
In this Letter, we proposed a novel method to optimize the
force-field parameters for protein systems.
Our method is knowledge-based in the sense that we
utilize the protein structures in the Protein Data Bank.
With this method one can optimize not only the existing force-field parameters
such as AMBER, CHARMM, OPLS, and so on, but also any force field as long
as its functional form is given.

We have presented an example of the application of the present method to
the original AMBER parm96 force field by optimizing the partial-charge
and backbone torsion-energy parameters.
There is still a lot of testing to be done.
For instance, other parameter sets in AMBER parm96 as well as other 
force fields should also be tried.
We used only 100 protein structures, and this number should be increased.
As for solvent effects in the optimization procedure, 
we employed the generalized-Born/surface area model,
but the validity of this approximation should be tested.  Namely, we have
to check how insensitive or sensitive the final optimized parameters are to
the solvation theory used.
Finally, because the parameter space we optimize may be large, we want to
use a more powerful optimization algorithm than simulated annealing.
Generalized-ensemble algorithm \cite{GEA} may be one choice of such optimization
methods.

\noindent
{\bf Acknowledgements}: \\
The computations were performed on the computers at the Research Center for
Computational Science, Okazaki National Research Institutes and at ITBL,
Japan Atomic Energy Research Institute.
This work was supported, in part, by 
the NAREGI Nanoscience Project, Ministry of Education, Culture, Sports, Science
and Technology, Japan.

% The Appendices part is started with the command \appendix;
% appendix sections are then done as normal sections
% \appendix

% \section{}
% \label{}

\newpage

\begin{table}
\caption{Partial-charge parameters of alanine}
\label{table-ala}
\vspace{0.3cm}
\begin{center}
\begin{tabular}{lrr} \hline
Atom &  parm96  &  optimized  \\ \hline
N    & $-0.4157$  &  $-0.3317$    \\
CA   &  0.0337  &   0.0628    \\
C    &  0.5973  &   0.4996    \\
HN   &  0.2719  &   0.2358    \\
O    & $-0.5679$  &  $-0.5533$    \\
HA   &  0.0823  &   0.0923    \\
CB   & $-0.1825$  &  $-0.0394$    \\
HB   &  0.0603  &   0.0112    \\ \hline
Total   & 0.0000  &  $-0.0003$   \\ \hline
\end{tabular}
\end{center}
\end{table}

%\newpage

\begin{table}
\caption{Partial-charge parameters of glutamic acid}
\label{table-glu}
\vspace{0.3cm}
\begin{center}
\begin{tabular}{lrr} \hline
Atom &  parm96  &  optimized  \\ \hline
N    & $-0.5163$  &  $-0.4207$    \\
CA   &  0.0397  &   0.0642    \\
C    &  0.5366  &   0.4635    \\
HN   &  0.2936  &   0.2618    \\
O    & $-0.5819$  &  $-0.6054$    \\
HA   &  0.1105  &   0.1255    \\
CB   &  0.0560  &   0.1215    \\
HB   & $-0.0173$  &  $-0.0387$    \\
CG   &  0.0136  &  $-0.0724$    \\
HG   & $-0.0425$  &  $-0.0307$    \\
CD   &  0.8054  &   0.8307    \\
OE   & $-0.8188$  &  $-0.8149$    \\ \hline
Total   & $-1.0000$  &  $-0.9999$   \\ \hline
\end{tabular}
\end{center}
\end{table}

%\newpage
\begin{table}
\caption{Partial-charge parameters of tyrosine}
\label{table-tyr}
\vspace{0.3cm}
\begin{center}
\begin{tabular}{lrr} \hline
Atom &  parm96  &  optimized  \\ \hline
N    & $-0.4157$  &  $-0.3492$    \\
CA   &  $-0.0014$  &   0.0437    \\
C    &  0.5973  &   0.5467    \\
HN   &  0.2719  &   0.2480    \\
O    & $-0.5679$  &  $-0.5705$    \\
HA   &  0.0876  &   0.0969    \\
CB   & $-0.0152$  &   0.0673    \\
HB   &  0.0295  &  $-0.0098$    \\
CG   & $-0.0011$  &  $-0.2136$    \\
CD   & $-0.1906$  &   0.0974    \\
HD   &  0.1699  &   0.0636    \\
CE   & $-0.2341$  &  $-0.3145$    \\
HE   &  0.1656  &   0.1376    \\
CZ   &  0.3226  &   0.2875    \\
OH   & $-0.5579$  &  $-0.5024$    \\
HH   &  0.3992  &   0.3968    \\ \hline
Total   & 0.0000  &  $-0.0002$   \\ \hline
\end{tabular}
\end{center}
\end{table}

%\newpage
\begin{table}
\caption{Backbone torsion-energy parameters}
\label{table-torsion}
\vspace{0.3cm}
%\begin{tabular}{lccc} \hline
\begin{center}
\begin{tabular}{lrrr} \hline
Dihedral angle     & $V_1/2$ & $V_2/2$ & $V_3/2$ \\ \hline
$\phi$ (parm96)    & 0.850   & 0.300   & 0.0   \\
$\phi$ (optimized) & 1.209   & 0.380   & $-0.018$  \\ \hline
$\psi$ (parm96)    & 0.850   & 0.300   & 0.0   \\
$\psi$ (optimized) & 0.133   & 0.578   & 0.050   \\ \hline
\end{tabular}
\end{center}
\end{table}

%\newpage

%\newpage

%\newpage

%\newpage

%\newpage

\end{document}